\documentclass[preprint]{aastex701}

\usepackage{multirow}
\usepackage{amsmath}

\usepackage{soul}
\usepackage{color}
\usepackage{cancel}

\begin{document}

\title{The Effects of Energy Conservation in Simulating Solar Eruptions}

\author[0009-0009-2176-6017]{Xianyu Liu} 
\affil{Department of Climate and Space Sciences and Engineering, University of Michigan, Ann Arbor, MI 48109, USA}
\email{xianyu@umich.edu}

\author[0000-0003-0176-4312]{Spiro K. Antiochos} 
\affil{Department of Climate and Space Sciences and Engineering, University of Michigan, Ann Arbor, MI 48109, USA}
\email{spiro.antiochos@gmail.com}

\author[0000-0001-9114-6133]{Nishtha Sachdeva} 
\affil{Department of Climate and Space Sciences and Engineering, University of Michigan, Ann Arbor, MI 48109, USA}
\email{nishthas@umich.edu}

\author[0000-0001-8459-2100]{G\'abor T\'oth} 
\affil{Department of Climate and Space Sciences and Engineering, University of Michigan, Ann Arbor, MI 48109, USA}
\email{gtoth@umich.edu}

\author[0000-0003-0472-9408]{Ward B. Manchester IV} 
\affil{Department of Climate and Space Sciences and Engineering, University of Michigan, Ann Arbor, MI 48109, USA}
\email{chipm@umich.edu}

\author[0000-0001-5260-3944]{Bart van der Holst} 
\affil{Astronomy Department, Boston University, Boston, MA 02215, USA}
\email{bartvand@bu.edu}

\author[0000-0002-6118-0469]{Igor V. Sokolov} 
\affil{Department of Climate and Space Sciences and Engineering, University of Michigan, Ann Arbor, MI 48109, USA}
\email{igorsok@umich.edu}

\author[0000-0001-9360-4951]{Tamas I. Gombosi} 
\affil{Department of Climate and Space Sciences and Engineering, University of Michigan, Ann Arbor, MI 48109, USA}
\email{tamas@umich.edu}

\author[0000-0003-3936-5288]{Lulu Zhao} 
\affil{Department of Climate and Space Sciences and Engineering, University of Michigan, Ann Arbor, MI 48109, USA}
\email{zhlulu@umich.edu}

\begin{abstract}

Strict energy conservation is, perhaps, the most basic principle in all physics, but has proven to be difficult to satisfy in numerical simulations of solar eruptions.  Here, the Alfv\'en Wave Solar atmosphere Model (AWSoM) is used to perform a rigorous comparison of CME simulations whose only difference is the use of a conservative vs. non-conservative scheme for the energy equation. A simple, symmetric active region is assumed for the initial magnetic field. As expected, the different numerical schemes result in very different plasma thermal energy, but surprisingly, we also find a factor $>2$ difference in the final kinetic energy, with the energy substantially larger in the energy-conservative scheme. The increase in thermal energy is comparable to the increase in kinetic energy in the conservative simulation. Our analysis reveals that the flare reconnection and increase of kinetic energy terminate earlier with the non-conservative scheme. We conclude that the plasma thermodynamics plays a critical role in the flare reconnection, with the thermal pressure gradient in the current sheet slowing down the reconnection. Our results imply that using strict energy-conservative numerics is critical for space weather modeling of CMEs and for understanding the CME energy budget partitioning.

\end{abstract}


\section{Introduction}\label{section_intro}

Coronal mass ejections (CME) are the most important form of solar activity that drive geomagnetic storms \citep[e.g.,][]{Gosling1993} and release enormous amounts of energy up to $>10^{33}$ erg\citep[e.g.,][]{Gopalswamy2010}. Consequently, the study of CME energetics has been attracting attention for many decades now. It is widely accepted that magnetic free energy in the corona is the major energy source for CMEs. This free energy is released mainly via flare reconnection \citep[][]{Forbes1991}. As summarized by \citet{Aschwanden2017}, flare reconnection involves (a) three primary processes: acceleration of non-thermal particles \citep[][]{Miller1997,Pesce-Rollins2024}, direct heating \citep[][]{Caspi2015}, and acceleration of CME ejeta, and (b) secondary processes: including chromospheric evaporation due to non-thermal particle precipitation \citep[][]{Antonucci1983,Antonucci1984} and particle acceleration driven by the CME shock \citep[][]{Kahler1978,Desai2016,Liu2025}. The energy of non-thermal particles will quickly be thermalized in the flare corona and chromosphere and contribute to thermal energy, resulting in the flare soft X-ray radiation burst, as in the well-known Neupert effect \citep[][]{Warmuth2020}. Consequently, the energy conversion can be viewed primarily as a transformation of free magnetic energy into bulk kinetic energy and plasma internal energy.

Calculating even this basic energy release process  in numerical models of CMEs/flares is challenging. The existing three-dimensional (3D) magnetohydrodynamic (MHD) simulations of CMEs do reproduce flare reconnection and the acceleration of the CME ejecta \citep[e.g.,][]{Karpen2012,Torok2018,Downs2021,Fan2017}, but they generally fail to capture the thermal energy. The problem is that the non-adiabatic plasma heating occurs mainly at the reconnecting current sheets, which are discontinuities on the MHD scale. As a result, the energy is lost via numerical diffusion unless special care is taken to explicitly conserve the energy numerically. This is especially difficult for a multi-temperature and anisotropic pressure plasma, conditions expected at discontinuities such as current sheets and shocks, because the standard pressure and temperature equations are not in explicit conservative form \citep{vanderHolst2025}. There have been, however, a few attempts to capture the thermal energy evolution in 3D CME simulations accurately. \cite{Reeves2019} described a single-fluid simulation where the numerically dissipated energy is accounted for in their energy diagnostics. Although not truly self-consistent, this procedure does conserve the total energy. CME simulations made with the BATS-R-US code \citep[e.g.,][]{Manchester:2004,Jin:2017a} have traditionally used an explicitly energy-conservative numerical scheme \citep[e.g.,][]{vanderholst:2014} to prevent the energy loss due to numerical diffusion. Energy conservation is particularly important to get the correct heating at CME-driven shocks, which has been addressed in prior work with BATS-R-US \citep{Manchester:2012,Jin:2013}. This treatment was also used in some recent works \citep[][]{vanderHolst2025,Wraback2026}. In \citet{vanderHolst2025}, they noted that the use of an energy-conservative scheme ensures global energetic consistency albeit the lack of microphysics in the current sheet, specifically non-thermal particle acceleration.

Despite these recent approaches for achieving energy conservation in CME simulations, it remains an open question as to whether energy conservation serves only to improve the thermodynamic consistency in the model, or whether it has significant effects on the dynamics of the eruption and magnetic field. The answer to this question is critical for understanding the energetics of CMEs and flares.  In this work, we will for the first time present a comparison between a pair of simulations of the same ideal CME event, one with the usual energy losses via numerical diffusion, and one with strict energy conservation numerically.  Section~\ref{section_method} describes our numerical schemes, physical treatments and simulation setup. The simulation results are given in Section~\ref{section_results}. We summarize our findings and conclusions in Section~\ref{section_summary}.

\section{Methods} \label{section_method}

\subsection{AWSoM Numerical Schemes}

Our simulations are based on the Alfve\'n Wave Solar atmosphere Model \citep[AWSoM, ][]{vanderholst:2014}. AWSoM uses the Block Adaptive Tree Solarwind-Roe-Upwind Scheme \citep[BATS-R-US, ][]{Powell1999} to solve the 3D MHD equations in a spherical domain extending from the bottom of the transition region at $r\approx1$ solar radius (hereafter $R_{s}$) to $24\, R_{s}$. The key functionality in AWSoM used in this work is the conservative and non-conservative schemes for the energy equation. Here we start from the conservative approach. AWSoM allows solving for the proton pressure ($P_{p}$) and electron pressure ($P_{e}$) using following equations \citep[e.g.,][]{vanderholst:2014}:
\begin{align}
&\frac{\partial}{\partial t}
\left(
    \frac{P_p}{\gamma - 1}
    + \frac{\rho u^2}{2}
    + \frac{B^2}{2\mu_0}
\right)
+ \nabla \cdot
\left[
    \left(
        \frac{\gamma P_p}{\gamma - 1}
        + \frac{\rho u^2}{2}
        + \frac{B^2}{\mu_0}
        + P_e
        + P_A
    \right)\mathbf{u}
    - \frac{\mathbf{B}(\mathbf{u}\cdot\mathbf{B})}{\mu_0}
\right]
- P_{e} \nabla\cdot\mathbf{u}
=\mathrm{S}_{p}
\label{eqn_pp}
\\
&\frac{\partial}{\partial t}
\left(
    \frac{P_e}{\gamma - 1}
\right)
+ \nabla \cdot
\left(
    \frac{P_e}{\gamma - 1}\mathbf{u}
\right)
+ P_e \nabla\cdot\mathbf{u}
=\mathrm{S}_{e}.
\label{eqn_pe}
\end{align}
Here $\rho$, $\mathbf{u}$, $\mathbf{B}$ indicate plasma density, bulk velocity, and magnetic field, respectively. The adiabatic index $\gamma$ is assumed to be $\frac{5}{3}$. $P_{A}$ is the Alfv\'en wave pressure, which will be described later. The source terms $\mathrm{S}_{p}$ and $\mathrm{S}_{e}$ are given by
\begin{align}
\mathrm{S}_{p}&=Q_p + P_{A} \nabla\cdot\mathbf{u} + \frac{N_p k_B}{\tau_{ep}}(T_e - T_p)
- \frac{GM_\odot \rho}{r^3}\mathbf{r}\cdot\mathbf{u},\label{eqn_Sp}\\
\mathrm{S}_{e}&=Q_e - \nabla\cdot\mathbf{q}_e
+ \frac{N_p k_B}{\tau_{ep}}(T_p - T_e) - Q_{\rm rad}.\label{eqn_Se}
\end{align}
Here $Q_{p,e}$ denotes the coronal heating terms for proton and electron. To account for the coronal heating process, AWSoM simulates the WKB transportation, non-WKB reflection, and phenomenological dissipation of Alfv\'en waves using two equations for energy densities of waves propagating parallel and anti-parallel to the magnetic field ($w_{\pm}$) \citep[][]{Chandran2011}:
\begin{equation}
\frac{\partial w_\pm}{\partial t}
+ \nabla \cdot \left[ \left( \mathbf{u} \pm \frac{\mathbf{B}}{\sqrt{\mu_{0}\rho}} \right) w_\pm \right]
+ \frac{w_\pm}{2} \left( \nabla \cdot \mathbf{u} \right)
=
\mp \mathcal{R} \sqrt{w_+ w_-}
- \Gamma_\pm w_\pm .
\end{equation}
Here $\mathcal{R}$ and $\Gamma_\pm$ denote the wave reflection and dissipation rates, respectively. The total coronal heating rate due to the Alfv\'en wave dissipation is $Q_{H}=\Gamma_{+}w_{+}+\Gamma_{-}w_{-}$. $Q_{p,e}$ are derived by partitioning $Q_{H}$ \citep[see Appendix B in ][]{vanderholst:2014}. Another effect of Alfv\'en wave is the WKB Alfv\'en wave pressure $P_{A}=\frac{w_{+}+w_{-}}{2}$, which appears in Equations~\ref{eqn_pp} and \ref{eqn_Sp}. The second term $-\nabla\cdot\mathbf{q}_e$ in Equation~\ref{eqn_Se} describes the electron heat conduction. The heat flux $q_{e}$ is given by the combination of the collisional Spitzer heat flux \citep[][]{Spitzer1953} and the collisionless heat flux \citep[][]{Hollweg1978}. The full expression of $q_{e}$ can be found in our previous works \citep[e.g.,][]{Shi2024,vanderholst:2014}. $\pm\frac{N_p k_B}{\tau_{ep}}(T_e - T_p)$ describes the heat exchange between protons and electrons through Coulomb collision, where $N_{p}$ denotes the proton and electron number density and $\tau_{ep}$ represents the relaxation time. The last term in Equation~\ref{eqn_Sp} describes the gravitational effect. $Q_{\rm rad}=Q_{\rm rad}=N_{e}N_{h}\Lambda(T_{e})$ denotes the optically thin radiative heat loss, where $N_{e}$ and $N_{h}$ are the electron number density and hydrogen number density, respectively. $\Lambda(T_{e})$ is the total radiative cooling function calculated from CHIANTI version 7.1 database \citep[][]{Landi2013}. 

\subsection{Conservative and Non-Conservative Energy Scheme}

Equations~\ref{eqn_pp} and \ref{eqn_pe} produce an energy-conservative numerical scheme. Since Equations~\ref{eqn_pp} and \ref{eqn_pe} use the same discretization for the spatial divergence, the two equations are discretely additive, giving
\begin{equation}
\frac{\partial e}{\partial t}
+ \nabla \cdot \mathbf{F}
=\mathrm{S}_{p}+\mathrm{S}_{e},
\label{eqn_e}
\end{equation}
where 
\begin{equation}
    e=\frac{P_p}{\gamma - 1}
    + \frac{P_e}{\gamma - 1}
    + \frac{\rho u^2}{2}
    + \frac{B^2}{2\mu_0}
\end{equation}
is the sum of the proton internal, electron internal, kinetic, and magnetic energy densities, and
\begin{equation}
    \mathbf{F}=\left(e+P_p+P_e+P_A\right)\mathbf{u} -\frac{\mathbf{B}(\mathbf{u}\cdot\mathbf{B})}{\mu_0}
\end{equation}
represents the energy flux. The conservative form of Equation~\ref{eqn_e} implies that Equations~\ref{eqn_pp} and \ref{eqn_pe}, together with the finite volume discretization in BATS-R-US, prevent the numerical dissipation of the sum of the four types of energy during the MHD numerical flux update. This feature is critical for capturing energy conversion at the flare current sheet and at any shocks in the system.

Let us now consider the numerical effects of the source terms in the equations above. Summing the two sources terms in Equation~\ref{eqn_e} yields:
\begin{equation}
    S_{p}+S_{e}=Q_{p} + Q_{e}
    + P_{A} \nabla\cdot\mathbf{u}
    - Q_{\rm rad}
    - \frac{GM_\odot \rho}{r^3}\mathbf{r}\cdot\mathbf{u}.
\end{equation}
Note that the form of the coronal heating term, radiative cooling term and gravitational term do not explicitly involve spatial derivative, indicating that these terms do not generate significant numerical diffusion near MHD discontinuities. For $P_{A} \nabla\cdot\mathbf{u}$, we find that $P_{A}$ is more than two orders of magnitude smaller than the thermal pressure and magnetic pressure in the flare current sheet of our simulations, which means $P_{A} \nabla\cdot\mathbf{u}$ is negligible compared to $\nabla \cdot \mathbf{F}$ there. We therefore conclude that the source terms do not violate energy conservation in the flare current sheets in our simulations. 

Equations~\ref{eqn_pp} and \ref{eqn_pe} have an analytically equivalent form \citep[e.g., ][]{toth2012adaptive}:
\begin{align}
    \frac{\partial P_{p}}{\partial t}+\nabla\cdot\left(P_{p}\mathbf{u}\right)+\left(\gamma-1\right)P_{p}\nabla\cdot\mathbf{u}&=\left(\gamma-1\right)
    \left[
    Q_p + 
    \frac{N_p k_B}{\tau_{ep}}(T_e - T_p)
    - \frac{GM_\odot \rho}{r^3}\mathbf{r}\cdot\mathbf{u}
    \right], \label{eqn_npp}
    \\
    \frac{\partial P_{e}}{\partial t}+\nabla\cdot\left(P_{e}\mathbf{u}\right)+\left(\gamma-1\right)P_{e}\nabla\cdot\mathbf{u}&=\left(\gamma-1\right)
    \left[
    Q_e + 
    \frac{N_p k_B}{\tau_{ep}}(T_p - T_e)
    - Q_{\rm rad}
    \right]. \label{eqn_npe}
\end{align}
Equations~\ref{eqn_npp} and \ref{eqn_npe} do not numerically guarantee energy conservation. AWSoM allows switching between the conservative scheme (Equations~\ref{eqn_pp} and \ref{eqn_pe}) and non-conservative scheme (Equations~\ref{eqn_npp} and \ref{eqn_npe}). We discuss below how the two types of schemes are used in our simulations.

\subsection{Simulation setup}

\begin{figure}[tp!]
\centering {\includegraphics[width=1.0\hsize]{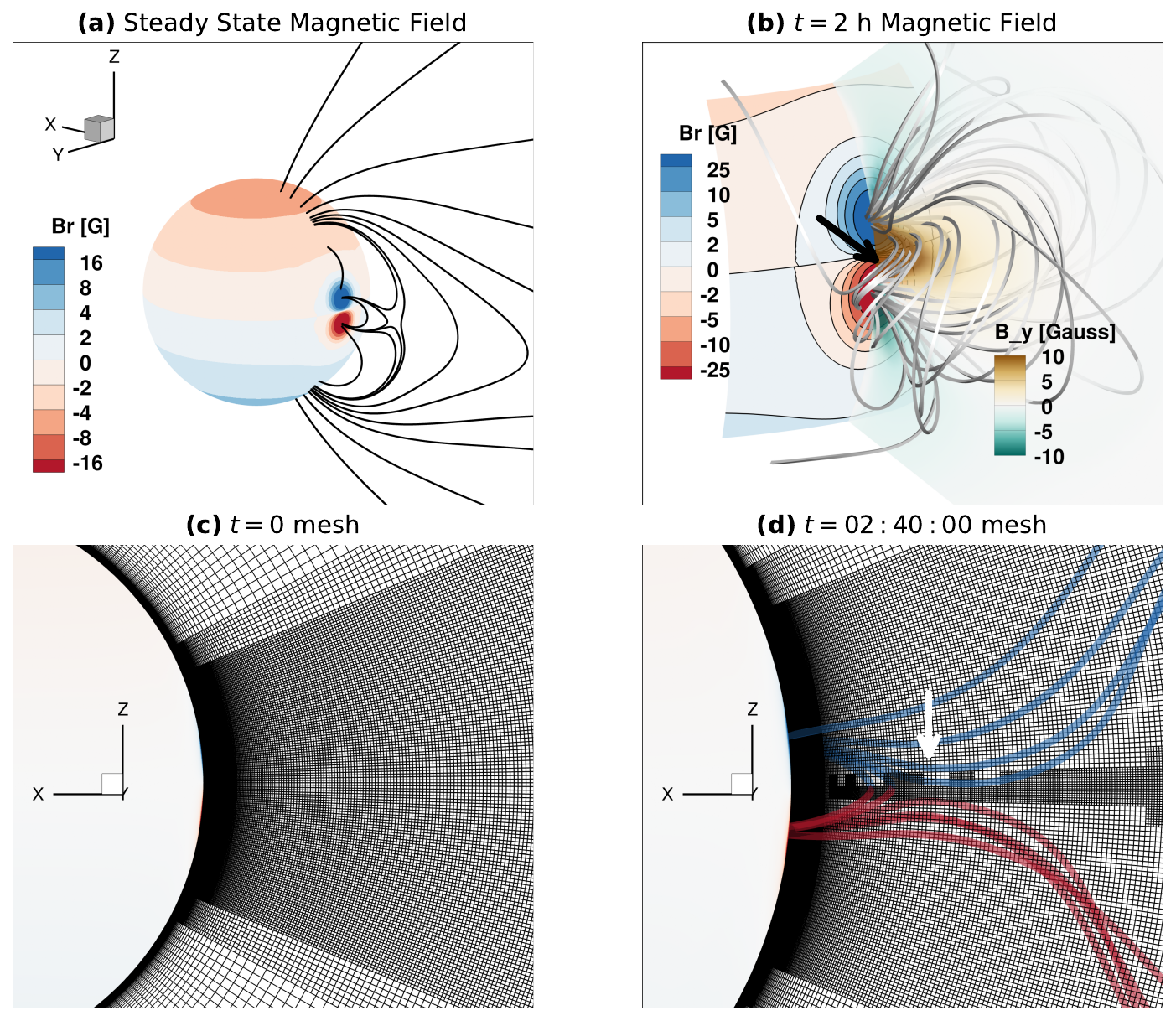}} 
\caption{Simulation setup. \textbf{(a)}: Magnetic field of the steady state background. \textbf{(b)}: Magnetic field lines and $B_{y}$ distribution in the $Y=0$ plane after the STITCH driving phase in Case N. The black arrow labels a sheared arcade produced by the driving. \textbf{(c)(d)}: Grid mesh in the $Y=0$ plane at $t=0$s and $t=9600$s, respectively. The color indicates the local sign of $B_{r}$ (blue for positive and red for negative) in magnetic field lines.}\label{fig_setup}
\end{figure}

Our simulations use the simple radial magnetic field ($B_{r}$) map shown in Figure~\ref{fig_setup}(a) as the boundary condition. This $B_{r}$ map is given by two pairs of subsurface magnetic monopoles:
\begin{equation}
    B_{r}\left(\theta,\phi\right)=\sum_{i=1,2}\sum_{\pm}\frac{\mu_{0}}{4\pi}\frac{\pm q_{i}^{m}}{\left|\mathbf{r}-\mathbf{r}_{i}\pm\mathbf{l}_{i}/2\right|^3}(\mathbf{r}-\mathbf{r}_{i}\pm\mathbf{l}_{i}/2)\cdot\mathbf{\hat{r}},
\end{equation}
where $i$ indicates the pair index and $\pm$ indicates two polarities. $\mathbf{r}_{i}$ denotes the position of center of the $i$th pair of monopoles. We chose $\mathbf{r}_{1}=(0,0,0)$ and $\mathbf{r}_{2}=(-0.95,0,0)R_{s}$ so that the two pairs of monopoles correspond to a global field and an active region field, respectively. $\mathbf{l}_{i}$ is the vector from the negative monopole to the positive monopole, and we chose $\mathbf{l}_{1}=(0,0,0.001)R_{s}$ and $\mathbf{l}_{2}=(0,0,0.2)R_{s}$. The magnetic charges $q_{1,2}^{m}$ are chosen so that: (1) the global dipole produces $\left|B_{r}\right|=5$ G at the poles, (2) the maximum $\left|B_{r}\right|$ in the active region is approximately $250$ G, and (3) the magnetic field exhibits a multi-polar configuration, which means the two pairs of monopoles are oppositely aligned. We use the $B_{r}$ map as the inner boundary condition for $B_{r}$.  The $B_{r}$ map is also used to determine the Poynting flux of the outgoing Alfv\'en waves ($S_{A}$) injected at the inner boundary:
\begin{equation}
    S_{A}=(S_\mathrm{A}/B)_{\odot}B_{r},
\end{equation}
where $(S_\mathrm{A}/B)_{\odot}=0.4$ erg s$^{-1}$ Mx$^{-1}$. 

We adopt specific thermodynamic treatments at or near the inner boundary. It is well-known that flare reconnection will lead to chromospheric evaporation \citep[e.g.][]{Antiochos1978}; therefore, although our inner boundary is at the top of the chromosphere, we use a relatively high number density of $2\times10^{18}$ m$^{-3}$ there. This assumption provides a plasma reservoir and allows the evaporation of plasma into the corona when a strong heat flux appears during the flare. The temperature at the inner boundary is $5\times10^{4}$ K. In regions where $T_{e}<2.2\times10^{5}$ K, we apply an artificial extension of the transition region \citep[e.g.,][]{Lionello2009,Sokolov2013,vanderholst:2014,Shi2024} to resolve the steep temperature gradient.

The CME simulations in this work use a steady-state solution of the corona, similar to those in previous works \citep[e.g.,][]{Sachdeva2019,Sachdeva2021,Shi2022,Liu2026eclipse}. We first obtained the 3D potential field source surface model \citep[PFSSM, ][]{Schatten1969,Toth2011} of the $B_{r}$ map with the source surface at $2.5R_{s}$ as the initial magnetic field of the steady-state simulation. An exponentially stratified atmosphere near the inner boundary, connected to the Parker solar wind solution \citep[][]{Parker:1958} farther out, is used to set the initial temperature, density, and velocity. The steady-state simulation uses the second-order shock-capturing Linde scheme with the third-order Koren limiter \citep[][]{Koren:1993} and performs 80,000 iterations with local time stepping. We applied adaptive mesh refinement \cite[AMR,][]{gombosi2003adaptive, toth2012adaptive} to the grid mesh and decomposed the domain into blocks of $6\times4\times4$ cells each. The simulation begins with an angular resolution of $2.8^{\circ}$ above the sphere $r=1.7R_{s}$ and a finer resolution of $1.4^{\circ}$ below. We applied two mesh refinements after 30,000 and 60,000 iterations to the region $(r,\phi,\theta)\in[1.00; R_s,5.0; R_s]\times[160^{\circ},200^{\circ}]\times[70^{\circ},110^{\circ}]$, and the resulting angular resolution within this region is $0.35^{\circ}$. Figure \ref{fig_setup}(a) shows the magnetic field of the steady-state solution. The magnetic field contains a null point at $(-1.62,0,0)\,R_{s}$ due to the multi-polar $B_{r}$ map

After obtaining the steady state solution, we turned on global time stepping and denote this time by $t=0$. We apply a dynamic AMR with a cadence of $60$ s to resolve the current sheets. The AMR detects the grid blocks where opposite signs of $B_{r}$ appear and refines these blocks to achieve an angular resolution of $0.18^{\circ}$. This refinement criterion is highly effective for capturing the flare current sheet. For those previously refined blocks that currently do not exhibit opposite signs of $B_{r}$ due to the magnetic field evolution, AMR applies a coarsening and reduces the resolution to $0.35^{\circ}$. We constrained this AMR within the region $(r,\phi,\theta)\in[1.1\; R_s,5.0\; R_s]\times[160^{\circ},200^{\circ}]\times[70^{\circ},110^{\circ}]$ to avoid an overwhelming number of cells. Figures~\ref{fig_setup} (c) and (d) show the grid mesh in the $Y=0$ plane for $t=0$ and $t=160$ min, well into the eruption. Note the high refinement is along the equatorial plane, exactly where the flare current sheet is located.

To drive a CME eruption, we adopted the STatistical Injection of Condensed Helicity Model \citep[STITCH, ][]{Dahlin2022,vanderHolst2025}, which models statistically the condensation of magnetic helicity due to the convective motions at the solar surface \citep[][]{Antiochos2013}. In our simulation, STITCH applies a source term to the horizontal magnetic field at $r=1.01R_{s}$:
\begin{equation}
    \frac{\delta \mathbf{B}_{s}}{\delta t}=\nabla_{s}\times\nabla_{r}\left(\zeta B_{r}\right) \label{stitch_formula},
\end{equation}
where $\zeta$ is modulated by the position. We apply STITCH within the region $(\phi,\theta)\in[160^{\circ},200^{\circ}]\times[70^{\circ},110^{\circ}]$. $\zeta$ peaks at $0.32\times10^{6}$ km$^{2}$s$^{-1}$ near the center of the active region and vanishes at the edges of this region. The STITCH driving is applied during the first $2$ hours of the simulation. The STITCH driving is applied during the first 2 hours of the simulation and the eruption begins shortly after. Figure~\ref{fig_setup} (b) presents the magnetic field at $t=2$ h in one simulation (Case N defined below) using the non-conservative energy scheme. The driving has produced a highly sheared arcade above the polarity inversion line (PIL) at the equator with a strong $B_{y}$ component. After $t=2$ h, the magnetic field evolution is fully self-consistent and results in an eruption even though the driving has been turned off. 

Our first comparison between conservative and non-conservative energy schemes focuses on the energy buildup phase. The simulation is branched into two cases: Case C, in which the conservative scheme described by Equations~\ref{eqn_pp} and \ref{eqn_pe} is enabled from $t=0$, and Case N, in which we use the non-conservative scheme described by Equations~\ref{eqn_npp} and \ref{eqn_npe} during the driving phase. Our second comparison focuses on the eruption after the driving. At $t=2$ h, we branch Case N into two sub-cases, Case NC and Case NN, depending on whether the conservative scheme is enabled or the non-conservative scheme is maintained. The reason for choosing Case N for the second branching will be given in the next section. Note that the choice of numerical scheme is the only setup difference among these cases. All the other parameters and settings are identical among all cases. Note also that for all the conservative runs the non-adiabatic heating is assumed to go only into the protons, but then it transfers to the electrons via collisions. The reason for this assumption is that Parker Probe direct measurements of heliospheric current sheet reconnection \citep{Eastwood2026}, laboratory plasma experiments \citep{Ono2011}, and kinetic simulations \citep{Drake2009} all show that ions gain an order of magnitude or more thermal energy than electrons during reconnection. Future studies will determine the effect of varying the ratio of proton to electron heating. 

\section{results} \label{section_results}

\begin{figure}[tp!]
\centering {\includegraphics[width=1.0\hsize]{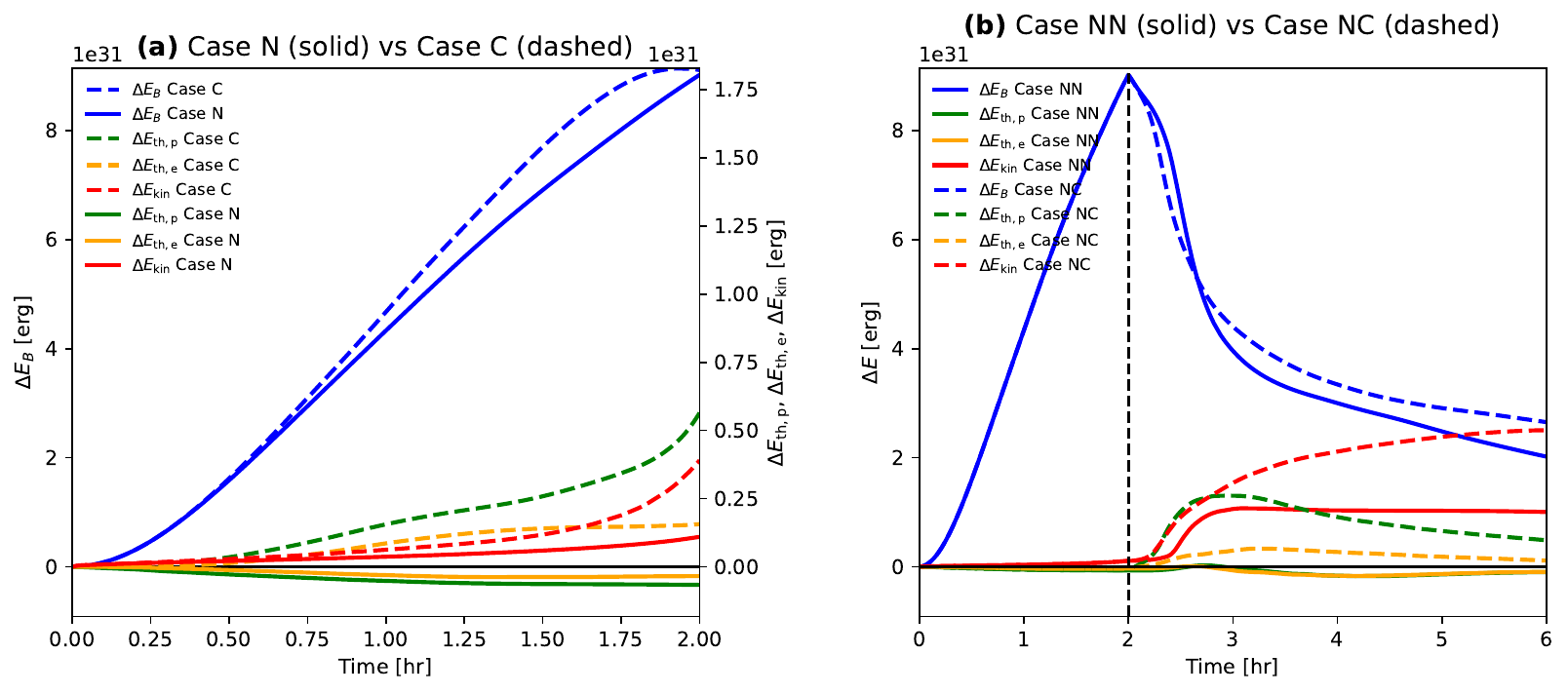}} 
\caption{Comparing the energy evolution of simulations using conservative and non-conservative schemes. The blue, red, green, and orange curves represent $\delta E_{B}$, $\delta E_{kin}$, $\delta E_{th,p}$, and $\delta E_{th,e}$. We use solid (non-conservative) and dashed (conservative) curves for distinguishing. \textbf{(a)}: Case C versus Case N. \textbf{(b)}: Case NC versus Case NN. }\label{fig_energy}
\end{figure}

Figure~\ref{fig_energy} shows the evolution of total magnetic energy ($E_{B}=\int\frac{B^{2}}{2\mu_{0}}dV$), kinetic energy ($E_{kin}=\int\frac{\rho u^{2}}{2}dV$), proton thermal energy ($E_{th,p}=\int\frac{P_{p}}{\gamma-1}dV$), and electron thermal energy ($E_{th,e}=\int\frac{P_{e}}{\gamma-1}dV$). The curves indicate the energy subtracting the initial value at $t=0$ ($\Delta E(t)=E(t)-E(0)$). In Figure~\ref{fig_energy} (a), the $\Delta E_{B}$ curves of Cases C and N reveal a relatively small difference $<10\%$. This means that the two cases give a similar amount of accumulated magnetic energy during the driving phase. The range for $\Delta E_{kin}$, $\Delta E_{th,p}$ and $\Delta E_{th,e}$ is only $20\%$ of that of $\Delta E_{B}$ in Figure~\ref{fig_energy} (a), showing that the change of thermal and kinetic energy is much smaller than $\Delta E_{B}$ in both cases. Nevertheless, $\Delta E_{kin}$, $\Delta E_{th,p}$, and $\Delta E_{th,e}$ in Case C become significantly larger than those in Case N. The reason is that in Case C, although the magnetic energy is slowly dissipated, it is strictly converted to the kinetic energy and thermal energy. The dissipated magnetic energy in Case N is lost from the system and does not show up as thermal energy or kinetic energy resulting from plasma expansion due to the heating. Our reason for using Case N as the starting point for simulating the ensuing eruption is the smaller kinetic energy in Case N compared to Case C. Observations show that the CME pre-eruptive configurations usually experiences a quasi-static state for days to weeks \citep[see the references in ][]{Chen2011}. Although Case N is less consistent in energetics, it better mimics the pre-eruptive quasi-static stage than Case C. The use of Case N also limit the effect of the CME speed gained during the driving.

In Figure~\ref{fig_energy} (b), the comparison of the CME eruption stage exhibits remarkable differences. First, the final $\Delta E_{kin}$ of Case NC ($\approx 2.50\times10^{31}$ erg) is $232\%$ of that in Case NN ($\approx 1.08\times10^{31}$ erg). One may argue that this difference can be due to the different amounts of released $E_{B}$ between Cases NC and NN. However, by calculating $\Delta E_{B}(t=6\mathrm{h})-\Delta E_{B}(t=2\mathrm{h})$, we found that the released magnetic energy in Case NC ($\approx 6.37\times10^{31}$ erg) is even smaller than that in Case NN ($\approx 7.00\times10^{31}$ erg), which implies a larger difference in the energy conversion rate. An interesting feature is that in Case NC, the rise of $E_{kin}$ continues till the end of the simulation, while in Case NN, $E_{kin}$ saturates at $t\approx 3$ h. This suggests that the relatively small $\Delta E_{kin}$ in Case NN is likely due to the termination of an energy conversion process. Another remarkable feature is that while both cases show a rapid increase phase in $\Delta E_{kin}$, the phase in Case NC occurs $\approx 10$ min earlier than in Case NN.

The second major difference in Figure~\ref{fig_energy} (b) is in the evolution of the thermal energy. Case NC generally shows much larger $\Delta E_{th,p}$ and $\Delta E_{th,e}$ than those in Case NN, which is consistent with the comparison of the energy buildup phase in panel (a). We notice that in Case NC during $t=2$ h$-3$ h, the proton thermal energy increases at a rate comparable to that of the kinetic energy. Although $E_{th,p}$ peaks at $t\approx 3$ h due to the radiative loss and proton-electron heat exchange, we notice that $\Delta E_{th,p}$ and $\Delta E_{kin}$ overall have the same order of magnitude during the eruption impulsive phase. This is consistent with observational results, which generally find that the total thermal energy gained during a CME/flare is comparable to the kinetic energy \citep[e.g.,][]{Ciaravella2001,Akmal2001,Emslie2004}.

\begin{figure}[tp!]
\centering {\includegraphics[width=1.0\hsize]{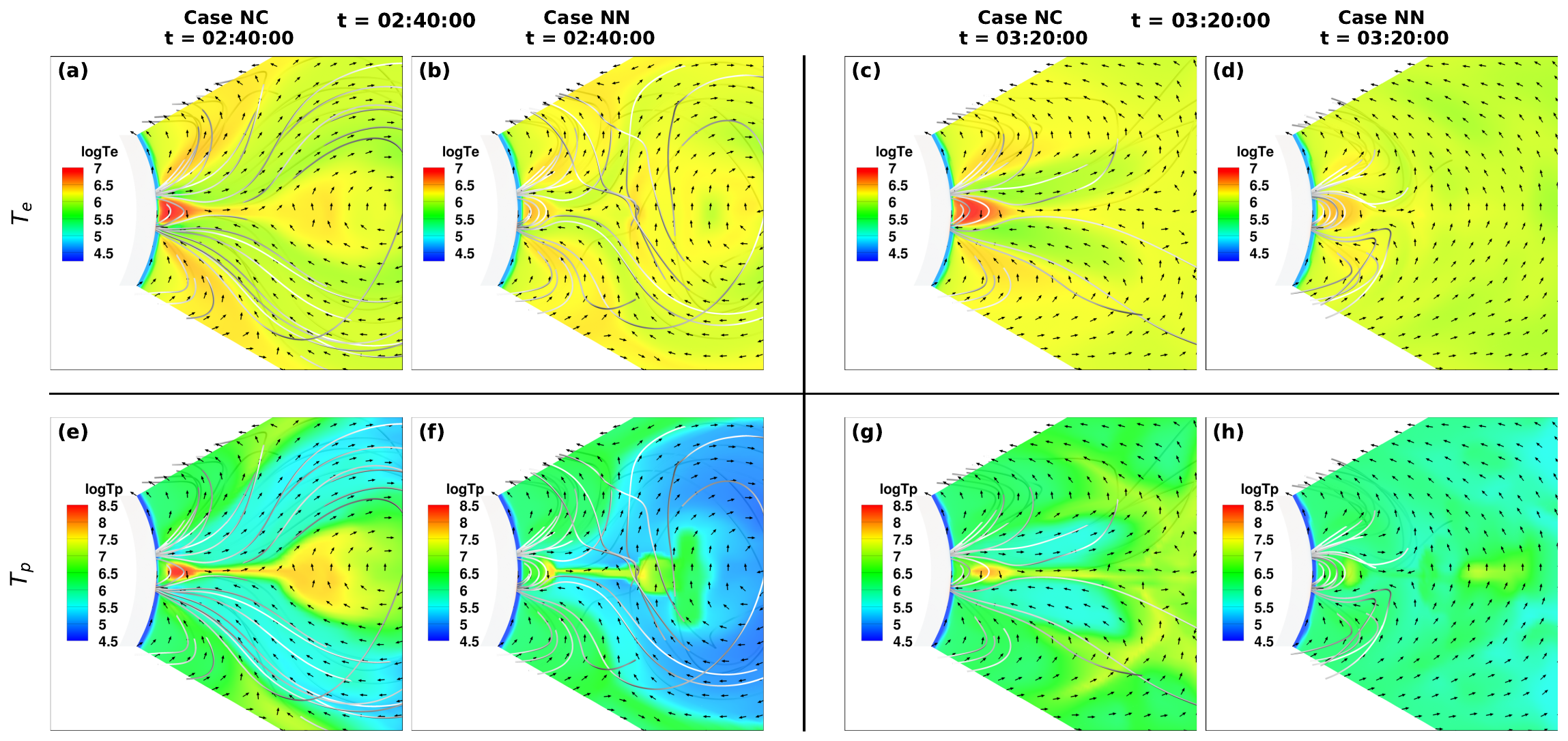}}
\caption{Temperature comparison between Cases NC and NN. \textbf{(a) (b)}: Electron temperature distribution in the $Y=0$ plane of Cases NC and NN at $t=02:40:00$. \textbf{(c) (d)}: Similar to \textbf{(a) (b)} but at $t=03:20:00$. The second row is similar to the first row, but for the proton temperature distribution. The black arrows indicate the direction of the tangential magnetic field vector in the $Y=0$ plane. The magnetic field lines are given in each panel for reference.}\label{fig_temperature}
\end{figure}

\begin{figure}[tp!]
\centering {\includegraphics[width=1.0\hsize]{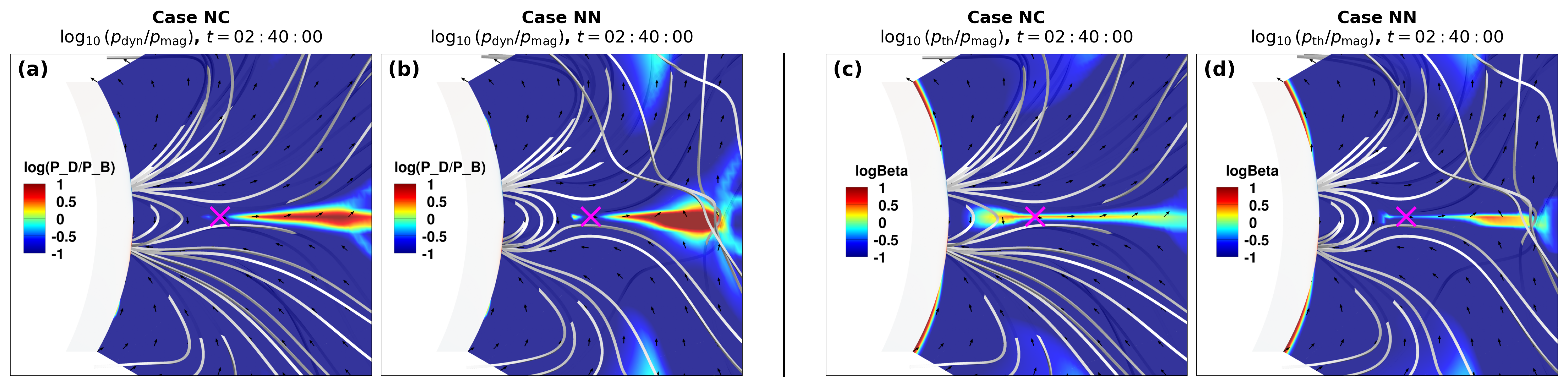}} 
\caption{\textbf{(a) (b)}: $P_{D}/P_{B}$ ratio for Cases NC and NN at $t=02:40:00$. \textbf{(c) (d)}: Similar to \textbf{(a) (b)} but for $P_{th}/P_{B}$. The magenta cross in each panel labels the reconnection site. }\label{fig_pressure_ratios}
\end{figure}

Besides the two major difference between Cases NC and NN described above, we also notice that for Case NC, $\Delta E_{kin}$ and $\Delta E_{th,p}$ respond differently to the switch to the energy-conservative scheme at $t=2$ h: the evolution $E_{kin}$ is relatively smooth at $t=2$ h, while $E_{th,p}$ gains a large time derivative as soon as the energy-conservative scheme is enabled. The reason is that enabling the energy-conservative scheme changes the method for calculating pressure, while the calculation of density and velocity is not immediately affected. This result shows that, as expected, the difference in thermodynamics between Cases NC and NN appears immediately as a primary effect, whereas the difference in kinetic energy is rather a secondary effect.

To understand the underlying physical reasons for our results, we present a comparison of the magnetic field and temperature distributions of Cases NC and NN in Figure~\ref{fig_temperature}. The first two columns show the comparison at $t=02:40:00$, which is during the early stage of the eruption. A flare current sheet with enhanced $T_{e}$ and $T_{p}$ is identifiable in either case. Generally, $T_{e}$ and $T_{p}$ in the current sheet and post-flare loops of Case NC are higher than those in Case NN by one order of magnitude, which is consistent with the higher thermal energy of Case NC. The electron temperature $T_{e}\approx10^{7}$ K in Case NC agrees with results from X-ray observations \citep[e.g., ][]{Hanneman2014,McKenzie1999}. The proton temperature in the current sheet in Case NC ($\approx10^{8}$K) is consistent with the prediction in \citet{Russell2025} that $T_{p}$ can exceed $60$MK. We also notice that the proton temperature of the post-flare loop in Case NC reaches $\approx 10^{8.5}$K. Since our simulation does not include non-thermal particles and assumes that all the non-adiabatic heating goes into the protons, which cool slowly via thermal conduction, the plasma evaporation from the inner boundary has a longer response time than usually calculated, leading to an underestimated density and an overestimated ion temperature at the top of the post-flare loop. Note, however, that the ion-electron energy transfer via collisions and the electron heat conduction are fully rigorous in our simulations. Figure~\ref{fig_temperature} (g) shows that at $t=$ 03:20:00, $T_{p}$ in the post-flare loops has already dropped below $\approx10^{8}$K. Note that the electron temperature is less affected since the electron heat conduction quickly transports the electron thermal energy away \citep[e.g.,][]{Yokoyama1997}.

A remarkable feature in the comparison for $t=$ 03:20:00, shown by the last two columns of Figure~\ref{fig_temperature}, is that the current sheet is still present in Case NC, while in Case NN the current sheet has disappeared. We examined the data and found that the current sheet in Case NC is preserved till the end of the simulation at $t=6$ h. This comparison suggests that Case NN has a much shorter duration of flare reconnection than Case NC. The consequence is that the flare reconnection in Case NC provides a long-lasting Lorentz force that accelerates the CME ejecta, while in Case NN the acceleration terminates early, which can explain the saturation of $E_{kin}$ in Case NN shown by Figure~\ref{fig_energy} (b).

To investigate the cause for the difference in reconnection, we calculate the ratio of dynamic pressure ($P_{D}=\rho u^{2}$) to magnetic pressure ($P_{B}=\frac{B^{2}}{2\mu_{0}}$) and the ratio of thermal pressure ($P_{th}=P_{p}+P_{e}$) to $P_{B}$, i.e., plasma $\beta$, at $t=$ 02:40:00 in Figure~\ref{fig_pressure_ratios}. In panels (a) and (b), the regions where $P_{D}/P_{B}$ exhibits enhancement indicate the reconnection outflow. In either case, we use the magenta cross between the up-flow and down-flow to label the reconnection site. In Figure~\ref{fig_pressure_ratios} (c) and (d), we notice a significant difference in plasma $\beta$. In Case NC, $\beta$ in the current sheet near the reconnection site is of order $10^{0.5-1}$, while in Case NN, $\beta$ is below $10^{0}$ near the reconnection site. The high $\beta$ in Case NC indicates that thermal pressure dominates over the magnetic pressure within the current sheet, which results in a strong thermal pressure gradient force that hinders the plasma inflow and reconnection. Comparatively, this effect is much weaker in Case NN due to the small $\beta$ value. The result is that Case NN has a higher reconnection rate, which explains the early termination of flare reconnection shown by Figure~\ref{fig_temperature}.

\section{Summary and Conclusions} \label{section_summary}

This paper presents 3D MHD simulations of an idealized flare/CME with conservative and non-conservative numerical schemes for the energy equation. We performed three simulations, labeled Case C, Case NC, and Case NN, for comparison of both the energy buildup and eruption phases. The energy buildup phase exhibits only minor differences between the two numerical schemes, with the conservative scheme producing a slight pre-eruptive increase of the kinetic energy and thermal energy. We expect that if the driving had been slower, the differences between the two schemes would be even smaller. On the other hand, we found several major differences in the CME eruption and flare reconnection between the conservative and non-conservative calculations, and investigated the physical processes responsible for the differences.

Our dramatic finding is that switching from the non-conservative scheme to the mathematically equivalent conservative scheme, while keeping all other settings identical, results in a difference in kinetic energy of over a factor of $2$. Of course, the factor $2.32$ applies only to our particular simulations and may vary with different settings, e.g., $B_{r}$ map, driving methods, and the grid resolution. Note, however, that flare reconnection in our simulation appears as an MHD discontinuity regardless of the resolution used, indicating that the difference between two schemes does not converge to zero with increasing resolution. Note also that our grid resolution ($\Delta \theta\approx0.18^{\circ}$) is comparable to those used in existing high-resolution CME simulations \citep[e.g.,][]{Torok2018,Reeves2019}. 

Our results also show that thermodynamics plays a critical role in flare reconnection. Our analysis of the flare reconnection implies that the thermal pressure within the flare current sheet impedes reconnection, which can help accelerate the CME. A weak thermal pressure gradient across the sheet can lead to fast reconnection and early termination of the conversion to kinetic energy, while a relatively strong thermal pressure gradient can result in a long-duration increase of CME kinetic energy. This thermodynamic difference appears to be the primary effect that distinguishes conservative and non-conservative simulations. The difference in the CME dynamics is likely a secondary effect due to the cumulative effect of thermodynamics in the reconnection region. These results may seem counterintuitive in that the use of a conservative scheme, which increases the plasma heating, also results in more mass acceleration, but in fact, it simply emphasizes the unphysical nature of the non-conservative calculation. In both simulations the flare currents collapse to a reconnecting sheet, but in the non-conservative case most of the energy released at the sheet simply disappears due to numerical diffusion. There is no feedback from the reconnection energy release back onto the dynamics as in a true physical system. This ``disappearance" drains the system of its free energy far too efficiently so that even the kinetic energy ends up reduced. Note that the magnetic energy of case NN drops well below that of case NC in Fig.~\ref{fig_energy} (b), showing the effect of this unphysical ``cooling". 

Another important feature of the energy conservation is that it seems to affect the current sheet formation process, as well as the current sheet relaxation by reconnection. We note from Fig.~\ref{fig_energy} that the rapid increase in $E_{kin}$ due to the onset of fast flare reconnection starts $\approx 10$ min earlier in Case NC than in Case NN. We believe that this is due to the immediate rise in thermal energy for the case NC, as discussed above, which helps expand the field outward, and thereby, causes the current sheet to collapse to the reconnection scale earlier. The effect is small, however, so further detailed studies are required in order to understand whether and how thermodynamics affects the current sheet formation and eruption onset. In particular, it would be highly instructive to perform studies in which the driving is very slow so the conservative scheme can be used for the complete current sheet formation process. 

From a space weather forecasting perspective, the key implication of our work is that whether consistent energetics is ensured in the CME simulation introduces a significant difference in the predicted CME kinetic energy. The use of energy-conservative schemes or treatments such as that found in \cite{Reeves2019} and in BATS-R-US \citet{Powell:1999} and demonstrated in this paper can improve the physical consistency of eruption energetics and enhance the accuracy of space weather predictions \citep[e.g.,][]{Toth:2007,Manchester:2014b,Jin:2017b,ChenHF:2025, Sachdeva:2026}.  Energy conservation is particularly important for accurate simulation of shock accelerated energetic particles (SEPs) and their forecast \citep[e.g.,][]{Sokolov:2004,ZhaoL:2024,LiuW:2025}. Our work also raises important issues that merit further investigations. For the particular simulations above, we found that the energy released into the thermal plasma is roughly equal to that in the mass motions, at least, in the impulsive phase, but this may well depend on the pre-eruption magnetic field configuration. In future work, we plan to study how the CME energy budget depends on specific properties of the pre-eruptive configuration using energy-conservative simulations. Furthermore, it will be essential to perform event studies and determine how well the conservative simulations can match actual observations of the kinetic energy to thermal energy ratio.

\begin{acknowledgments}
This work is supported by the NASA LWS Strategic Capabilities grant 80NSSC22K0892 (SCEPTER). We thank Dr. Jon Linker for valuable discussions on this work. We acknowledge the high-performance computing support from the Texas Advanced Computing Center (TACC) Frontera at the University of Texas at Austin \citep{stanzione2020frontera}.
\end{acknowledgments}

\bibliography{ref,ref_CMEs}
\bibliographystyle{aasjournalv7}

\end{document}